\documentclass[runningheads]{llncs}

\usepackage{graphicx}
\usepackage{url}
\usepackage{multirow}
\usepackage{tabularx}

\begin{document}

\title{The Knowledge Graph Track at OAEI}
\subtitle{Gold Standards, Baselines, and the Golden Hammer Bias}
\author{Sven Hertling\orcidID{0000-0003-0333-5888} \and
Heiko Paulheim\orcidID{0000-0003-4386-8195}}

\authorrunning{S. Hertling and H. Paulheim}

\institute{University of Mannheim\\Data and Web Science Group\\B6 26, 68159 Mannheim\\\email{\{sven,heiko\}@informatik.uni-mannheim.de}}

\maketitle

\begin{abstract}
The Ontology Alignment Evaluation Initiative (OAEI) is an annual evaluation of ontology matching tools.
In 2018, we have started the Knowledge Graph track, whose goal is to evaluate the simultaneous matching of entities and schemas of large-scale knowledge graphs.
In this paper, we discuss the design of the track and two different strategies of gold standard creation.
We analyze results and experiences obtained in first editions of the track, and, by revealing a hidden task, we show that all tools submitted to the track (and probably also to other tracks) suffer from a bias which we name the \emph{golden hammer bias}.

\end{abstract}

\section{Introduction}
The Ontology Alignment Evaluation Initiative (OAEI)\footnote{\url{http://oaei.ontologymatching.org/}} was started in 2004 as a forum to collect benchmark datasets for ontology matching tools, and a regular evaluation of those tools \cite{euzenat2011ontology}. Over the years, new tracks with different foci have been added, e.g., for instance matching in 2009 \cite{ferrara2013evaluation}, for multi-lingual ontology matching in 2012 \cite{meilicke2012multifarm},  for interactive matching in 2013 \cite{paulheim2013towards}, and for the discovery of complex alignments in 2018 \cite{thieblin2018first}.

The general setup of OAEI tracks is that users can download pairs of input ontologies and have to provide the correspondences (in general: pairs of equivalent classes, properties, and/or instances). Up to 2009, participants in the challenge ran their tools on their own machines and submitted the results, which gave way to over-tuning to specific tasks (i.e., finding optimal parameter sets for individual tasks rather than developing tools that deliver decent results consistently across different tracks). 

From 2010 on, the format of OAEI was subsequently changed from the submission of \emph{results} to the submission of \emph{systems}, which where then run centrally by the organizers using the SEALS platform \cite{wrigley2012semantic}. This also gave way for controlled measurements of computational performance. Since 2012, all tracks of OAEI are conducted using the SEALS platform, since 2018, the HOBBIT platform is used as second platform next to SEALS \cite{jimenez2018introducing}.

In 2018, we introduced a new track, i.e., the Knowledge Graph track \cite{algergawy2018results}. Since most of the other tracks focused \emph{either} on schema or instance matching, the objective was to evaluate tools that solve both tasks in a real-world setting: as more and more knowledge graphs are developed, the discovery of links both on the instance and schema level becomes a crucial task in combining such knowledge graphs \cite{ringler2017one}.

The rest of this paper is structured as follows. Section~\ref{sec:data} describes the track, the datasets used and the two different strategies employed to create the gold standard for the 2018 and 2019 edition of the track. Section~\ref{sec:results} discusses the results from the 2019 edition, as well as the observation of the golden hammer bias in an additional evaluation. We close with a summary and an outlook on future work.

\section{Data for the Matching Tasks}
\label{sec:data}
The data for the knowledge graph matching track is taken from the DBkWik project \cite{hertling2018dbkwik,hertling2019dbkwik}. In that project, we execute the DBpedia Extraction Framework \cite{LehmannDBpedia} on several different Wikis from \emph{Fandom}\footnote{\url{http://www.fandom.com/}}, which is one of the most popular \emph{Wiki Farms}, comprising more than 385,000 individual Wikis totaling more than 50 million articles.
The result is a multitude of disconnected knowledge graphs, i.e., one knowledge graph extracted per Wiki, where each entity is derived from one page in a Wiki. In order to fuse those into one coherent knowledge graphs, we have to identify instance matches (i.e., entities derived from pages about the same real-world entity in different Wikis) as well as schema matches (i.e., classes and properties derived from different constructs in different Wikis).

\subsection{Knowledge Graphs}
For the 2018 and 2019 edition of the track, we picked groups of Wikis with a high topical overlap (RuneScape, Marvel comics, Star Trek, and Star Wars). Those are depicted in table~\ref{tab:KGtrack}. The groups cover different topics (movies, games, comics, and books)\footnote{More details are available at \url{http://oaei.ontologymatching.org/2019/knowledgegraph/index.html}}. 

Moreover, as a hidden evaluation task for the 2019 edition, we added one more Wiki which has almost \emph{no} topical overlap with the above, but a large likelihood of having many instances with the same name. To that end, we chose the Lyric Wiki, containing around 2M instances (mostly songs, albums, and artists). For example, there are multiple songs named \emph{Star Wars} and \emph{Star Trek}, which, however, should \emph{not} be matched to the movie or series of the same name.

\subsection{Gold Standard 2018}
For creating the gold standard for evaluation, we took a two-fold approach. The schema level (i.e., classes and properties) are mapped by experts.

\begin{figure}[t]
	\centering
	\includegraphics[width=\textwidth]{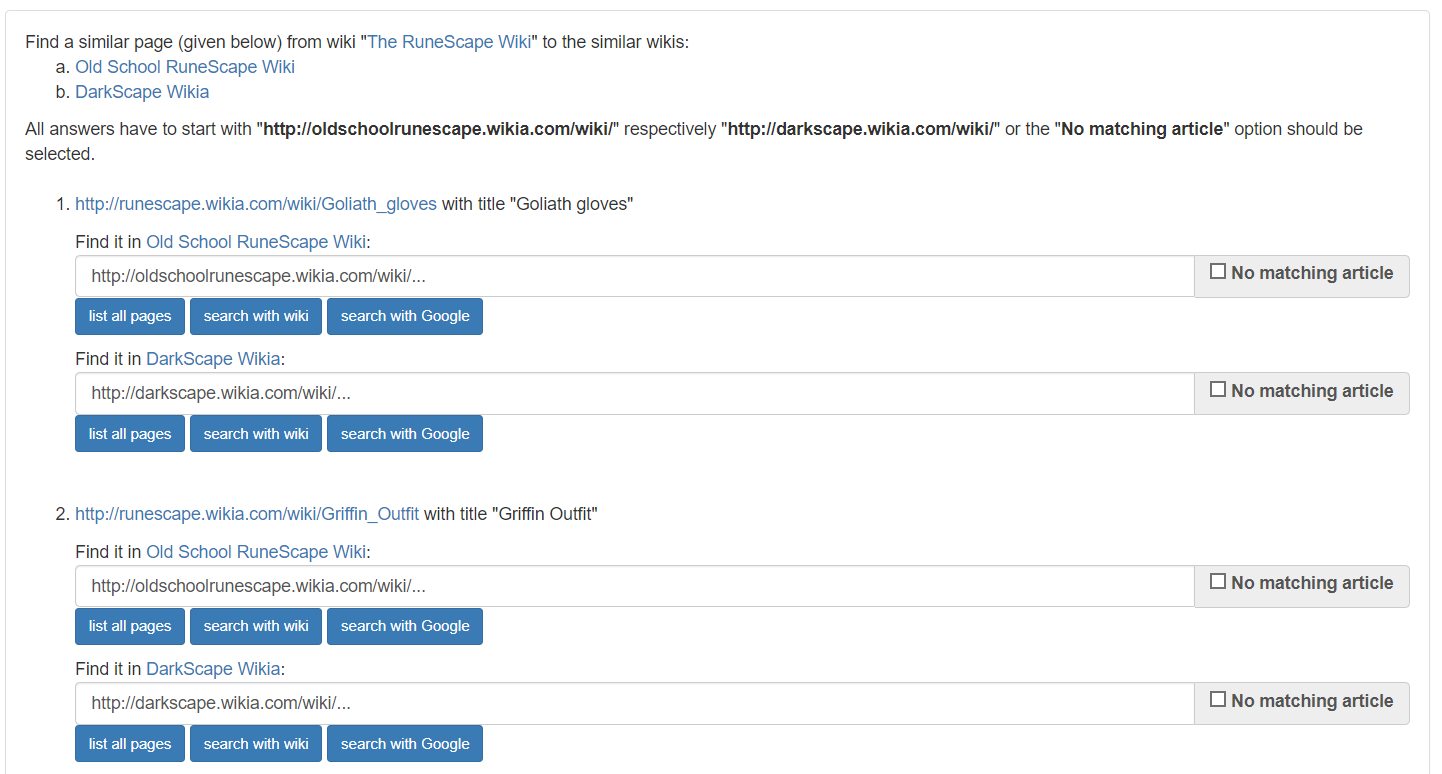}
	\caption{Crowd-sourcing interface.}
	\label{fig:crowdSourcingInterface}
\end{figure}

\begin{table}[t]
    \caption{Knowledge Graphs used in the 2018 and 2019 editions of the OAEI Knowledge Graph track. The numbers correspond to the 2019 version where applicable.}
    \label{tab:KGtrack}
	\centering
	\begin{tabularx}{\textwidth}{|X|l|r|r|r|c|c|}
		\hline
        Source Wiki & Hub  & Instances & Properties & Classes & 2018 & 2019 \\
        \hline
        RuneScape & Games  &  200,605 & 1,998 & 106 & X & \\
        Old School RuneScape & Games  &  38,563 & 488 & 53 & X &  \\
        DarkScape   & Games  &  19,623 & 686 & 65 & X &  \\
        \multicolumn{7}{|l|}{\textit{Characteristic classes: item, bonus, non-player character, recipe, monster, music }}\\
        \hline
        Marvel Database  & Comics  &  210,996 & 139 & 186 & X & X \\
        Hey Kids Comics & Comics  &  158,234 & 1,925 & 181 & X &  \\
        DC Database  & Comics  &  128,495 & 177 & 5 & X &  \\
        Marvel Cinematic Universe (mcu) & Movies  & 17,187 & 147 & 55 &  & X \\
        \multicolumn{7}{|l|}{\textit{Characteristic classes: actor, character, filmmaker, location, music, episode, event}}\\
        \hline
        Memory Alpha  & TV  & 45,828 & 325 & 181 & X & X \\
        Star Trek Expanded Universe  & TV  &  13,426 & 202 & 283 & X & X \\
        Memory Beta  & Books  &  51,323 & 423 & 240 & X & X \\
        \multicolumn{7}{|l|}{\textit{Characteristic classes: actor, individual, character, starship, comic, planet, species}}\\
        \hline
        Star Wars & Movies  &  145,033 & 700 & 269 &  & X \\
        The Old Republic & Games  &  4,180 & 368 & 101 &  & X \\
        Star Wars Galaxies  & Games  & 9,634 & 148 & 67 &  & X \\
        \multicolumn{7}{|l|}{\textit{Characteristic classes: character, planet, species, battle, weapon, comic book, item}}\\
		\hline
		Lyrics & Music & 1,062,920 & 270 & 67 &  & X \\
		\multicolumn{7}{|l|}{\textit{Characteristic classes: song, album, artist, translation, collaboration}}\\
		\hline
	\end{tabularx}
\end{table}

For mapping the instance-level, we used a crowd-sourcing approach on Amazon MTurk. As shown in Fig.~\ref{fig:crowdSourcingInterface}, users were presented a page link to a Wiki page for one Wiki (these pages were randomly sampled), and asked to identify a matching page in two other Wikis. In order to ease the task, they were provided with links to the Wiki's search function and Google site search. Each task was evaluated by five crowdworkers, and we added mappings to our gold standard if the majority agreed on it. Since the task was to match an entity in one source Wiki to two target Wikis, we also add mappings \emph{between} the two target Wikis if the entity is matched to both. This setting was executed for 3 groups of Wikis sharing the same domain and each Wiki of each group was used once as a source. Overall, the inter annotator agreement was 0.87 (according to \cite{fleiss_interpretation}, this is an \emph{almost perfect agreement}). 

The result is a partial gold standard for nine pairs of knowledge graphs, as depicted in Table~\ref{tab:gold2018}. A special characteristic of this gold standard is that non-matching entities are also contained explicitly (i.e,. crowdworkers agreed that they could not find a matching page in the other Wiki).

From table~\ref{tab:gold2018}, it can be observed that the gold standard contains mostly trivial matches (92.6\% of the class matches, 82.4\% of the property matches, and 93.6\% of the instance matches) which is an exact string match of the label. One possible reason is that crowdworkers were probably not motivated to search for matching pages if they could not find them easily based on matching names, and the provision of search links to ease the task might have increased that bias. Another reason might be the random sampling of source pages. In most cases the page creators give the same name to a well-known concept and only a few pages have different titles. With the given sampling method, the probability to have such pages in the resulting sample is rather low.

\begin{table}[t]
    \caption{Size of the Gold Standard used for OAEI 2018. The numbers in parantheses also count the negative mappings.}
    \label{tab:gold2018}
    \centering
    \begin{tabularx}{\textwidth}{|X|r|r|r|r|r|r|}
        \hline
         & \multicolumn{2}{c|}{Class} & \multicolumn{2}{c|}{Property} & \multicolumn{2}{c|}{Instance} \\
         \hline
         &       & non-    &       & non- &          & non- \\
         & total & trivial & total & trivial & total & trivial \\
        \hline
        darkscape-oldschoolrunescape & 11 \hphantom{1}(18) & 2 & 14 \hphantom{1}(20) & 1 & 46 \hphantom{1}(84) & 2 \\
        runescape-darkscape & 15 \hphantom{1}(20) & 1 & 10 \hphantom{1}(20) & 0 & 73 \hphantom{1}(86) & 1 \\
        runescape-oldschoolrunescape & 13 \hphantom{1}(17) & 1 & 12 \hphantom{1}(20) & 1 & 51 \hphantom{1}(88) & 4 \\
        \hline
        heykidscomics-dc & 2 \hphantom{1}(15) & 0 & 10 \hphantom{1}(20) & 2 & 25 \hphantom{1}(78) & 4 \\
        marvel-dc & 2 \hphantom{11}(5) & 0 & 8 \hphantom{1}(20) & 1 & 7 \hphantom{1}(72) & 2 \\
        marvel-heykidscomics & 2 \hphantom{1}(12)& 0 & 10 \hphantom{1}(20) & 2 & 22 \hphantom{1}(64) & 1 \\
        \hline
        memoryalpha-memorybeta & 0 \hphantom{1}(11) & 0 & 10 \hphantom{1}(20) & 7 & 19 \hphantom{1}(68) & 0 \\
        memoryalpha-stexpanded & 0 \hphantom{11}(3) & 0 & 9 \hphantom{1}(20) & 1 & 9 \hphantom{1}(69) & 1 \\
        memorybeta-stexpanded & 0 \hphantom{1}(14) & 0 & 8 \hphantom{1}(20) & 1 & 12 \hphantom{1}(67) & 2 \\
        \hline
        Total & 54 (115) & 4 & 91 (180) & 16 & 264 (676) & 17 \\
        \hline
    \end{tabularx}
\end{table}

During OAEI 2018, five systems were evaluated on the KG track: AML \cite{faria2013agreementmakerlight}, POMap++ \cite{laadhar2018oaei}, Holontology \cite{roussille2018holontology}, DOME \cite{hertling2018dome}, and three variants of LogMap (LogMap, LogMapBio, LogMapLt) \cite{jimenez2011logmap}. Additionally, we also used a string equivalence baseline. Due to the large number of trivial correspondences, none of the systems was able to beat the simple string equivalence baseline \cite{algergawy2018results}. 

\subsection{Gold Standard 2019}
For the 2019 edition of the knowledge graph track, we followed a different approach. While the schema level interlinks were still created by experts, we exploited explicit interlinks between Wikis for the instance level, pages in Wikis with links to a corresponding page in another Wiki. To that end, we selected five pairs of Wikis which have a large number of such interlinks.

Due to the fact that not all inter wiki links on a page link two pages aout the same entity, a few restrictions were made: 1) Only links in sections with a header containing \emph{link} are used e.g. as in ``External links''\footnote{an example page with such a section is \url{https://memory-alpha.fandom.com/wiki/William_T._Riker}}, 2) all links are removed where the source page links to \emph{more than one} page in another wiki (ensures the alignments are functional), and 3) multiple links which point to the same concept are also removed (ensures injectivity). The underlying assumption of the latter two steps is that in each wiki (similar to Wikipedia), only one page per entity (e.g., person, song, movie) exists.
As a preprocessing step, for each of those links, we executed an HTTP request to resolve potential redirects. Thus we always end up with the same identifier (URL) for one concept. Like the 2018 gold standard, this gold standard is only a \emph{partial gold standard}, but without any explicit negative mappings.

Table~\ref{tab:gold2019} shows the resulting gold standard. It can be observed that the fraction of non-trivial matches is considerably larger, especially on the instance level. Moreover, the absolute number of instance matches is also two magnitudes larger than in the 2018 gold standard.

\section{Results and Observations}
The two gold standards were used in the 2018 and 2019 editions of OAEI for a new knowledge graph track. Different tools were submitted to both editions, which allowed for a variety of insights.

In both years, the evaluation was executed on a virtual machine (VM) with 32GB of RAM and 16 vCPUs (2.4 GHz), with Debian 9 operating system and Openjdk version 1.8.0\_212, using the SEALS client (version 7.0.5). The alignments generated by the participating tools were evaluated based on precision, recall, and f-measure for classes, properties, and instances (each in isolation). Moreover, we report the overall precision, recall, and f-measure across all types.

As a baseline, we employed two simple string matching approaches. The source code for these baseline matchers is publicly available.\footnote{\url{http://oaei.ontologymatching.org/2019/results/knowledgegraph/kgBaselineMatchers.zip}}

\begin{table}[t]
    \caption{Size of the Gold Standard used for OAEI 2019}
    \label{tab:gold2019}
    \centering
    \begin{tabularx}{\textwidth}{|X|r|r|r|r|r|r|}
         \hline
         & \multicolumn{2}{c|}{Class Matches} & \multicolumn{2}{c|}{Property Matches} & \multicolumn{2}{c|}{Instance Matches} \\
         &       & non-    &       & non- &          & non- \\
         & total & trivial & total & trivial & total & trivial \\
        \hline
        starwars-swg & 5 & 2 & 20 & 0 & 1,096 & 528 \\
        starwars-swtor & 15 & 3 & 56 & 6 & 1,358 & 220 \\
        \hline
        mcu-marvel & 2 & 0 & 11 & 0 & 1,654 & 726 \\
        \hline
        memoryalpha-memorybeta & 14 & 10 & 53 & 4 & 9,296 & 2,140 \\
        memoryalpha-stexpanded & 13 & 6 & 41 & 3 & 1,725 & 274 \\
        \hline
        Total & 49 & 21 & 181 & 13 & 15,129 & 3,888 \\
        \hline
    \end{tabularx}
\end{table}

\label{sec:results}
\subsection{Results from OAEI 2018}
In 2018, only a simple string equivalence across normalized strings was used, whereas in 2019, we also incorporated similarities of the alternative labels (\texttt{skos: altLabel}) present in the knowledge graphs as a second baseline. These labels were generated by using all titles of redirect pages in the Wikis, and they often contain synonym or shorter versions of the original title. This should in general increase the recall but lower the precision of a matching approach. For example, \emph{Catarina} redirects to \emph{Kathryn Janeway} in the \emph{memoryalpha} Wiki\footnote{\url{https://memory-alpha.fandom.com/wiki/Special:WhatLinksHere/Kathryn_Janeway?hidelinks=1&hidetrans=1}}, so the baseline would consider all entities with the label \emph{Catarina} as matches for the entity \emph{Kathryn Janeway} derived from that Wiki.

The results for OAEI 2018 are depicted in table~\ref{tab:KGtrackResults2018}. Precision was computed based on the explicit negative mappings present in the 2018 gold standard. Four key observations can be made:
\begin{enumerate}
        \item Except for LogMap, LogMapLt, and AML, all participating systems could solve all tasks.
        \item The runtime varies greatly, ranging from five minutes to almost four hours for solving all nine tasks.
        \item Except for DOME, no matcher is capable of matching properties.
        \item Overall, the string baseline is hard to beat. Only two matchers (DOME and LogMapBio) outperform the baseline for classes, none for properties and instances.
\end{enumerate}
The first two observations show that in principle, existing ontology matching tools can actually match knowledge graphs, although with different computational behavior.

The third observation is due to a special characteristic of the underlying datasets. While standard ontology matching tools expect OWL Lite or DL ontologies, in which properties are properly typed as \texttt{owl:ObjectProperty} and \texttt{owl:Datatype\-Property}, the DBkWik knowledge graphs have a very shallow schema which does not make that distinction. Instead, all properties are marked as \texttt{rdf:Property}. 

The fourth observation may be attributed to the characteristics of the 2018 gold standard: as discussed above, a (probably overestimated) large fraction of matches is trivial, so that trivial matching approaches have an unrealistic advantage in this setup. This observation, together with the desire to have a larger-scale gold standard, lead to the new gold standard used in the 2019 edition.

\subsection{Results from OAEI 2019}
For the evaluation of the 2019 tracks, we did not have any explicit negative mappings. Hence, we exploited the fact that our partial gold standard contained only 1:1 correspondences, and we further assume that in each knowledge graph, only one representation of each entity exists (typically, a Wiki does not contain two pages about the same real-world entity). This means that if we have a correspondence $<a,b>$ in our gold standard, and a matcher produces a correspondence $<a,b'>$ to a different entity, we count that as a false positive. The count of false negatives is only increased if we have a 1:1 correspondence and it is not found by a matcher. The whole source code for generating the evaluation results is also available.\footnote{\url{http://oaei.ontologymatching.org/2019/results/knowledgegraph/matching-eval-trackspecific.zip}}

As a pre-evaluation check, we evaluated all SEALS participants in the OAEI (even those not registered for the track) on a very small matching task.\footnote{\url{http://oaei.ontologymatching.org/2019/results/knowledgegraph/small_test.zip}} This revealed that not all systems were able to handle the task, and in the end, only the following systems were evaluated: AGM~\cite{AGM19}, AML~\cite{AML19}, DOME~\cite{DOME19}, FCAMap-KG~\cite{FCAMap19}, LogMap~\cite{logmap19}, LogMapBio, LogMapKG, LogMapLt, POMap++~\cite{pomap19}, Wiktionary~\cite{wiktionary19}. Out of those, only LogMapBio, LogMapLt and POMap++ were not registered for this track. Holontology, which participated in 2018, did not submit a system to OAEI 2019. 

In comparison to 2018, more matchers participated and returned meaningful correspondences. Moreover, there are systems and system variations which especially focus on the knowledge graph track, e.g., FCAMap-KG and LogMapKG.
\begin{table}[t!] 
	\centering
	\caption{Knowledge graph track results for 2018, divided into class, property, instance, and overall correspondences. \cite{algergawy2018results}}
	\begin{tabular}{|l|c|c|c|c|c|c|}
		\hline
		System & Time & \# tasks & Size & Prec. & F-m. & Rec. \\
		\hline
		\multicolumn{7}{|c|}{Class performance}\\
		\hline
		AML&24:34:08&5&11.6&0.85 (0.85)&0.64 (0.87)&0.51 (0.88)\\
		POMAP++&0:07:18&9&15.1&0.79 &0.74 &0.69 \\
		Holontology&0:05:18&9&16.8&0.80&0.83 &0.87 \\
		DOME&3:49:07&9&16.0&0.73 &0.73&0.73  \\
		LogMap&3:54:43&7&21.7&0.66 (0.66)&0.77 (0.80)&0.91 (1.00)\\
		LogMapBio&0:39:00&9&22.1&0.68&0.81 &1.00 \\
		LogMapLt&0:08:20&6&22.0&0.61 (0.61)&0.72 (0.76)&0.87 (1.00)\\
		Baseline&0:06:52&9&18.9&0.75 &0.79 &0.84\\
		\hline
		\multicolumn{7}{|c|}{Property performance}\\
		\hline
		AML&24:34:08&5& 0.0&0.0 &0.0 &0.0 \\
		POMAP++&0:07:18&9& 0.0&0.0 &0.0 &0.0 \\
		Holontology&0:05:18&9& 0.0&0.0 &0.0 &0.0 \\
		DOME&3:49:07&9& 207.3&0.86 &0.84 &0.81 \\
		LogMap&3:54:43&7& 0.0&0.0 &0.0&0.0 \\
		LogMapBio&0:39:00&9& 0.0&0.0 &0.0 &0.0 \\
		LogMapLt&0:08:20&6& 0.0&0.0 &0.0 &0.0 \\
		Baseline&0:06:52&9& 213.8&0.86 &0.84 &0.82\\
		\hline
		\multicolumn{7}{|c|}{Instance performance}\\
		\hline
		AML&24:34:08&5& 82380.9&0.16 (0.16)&0.23 (0.26)&0.38 (0.63)\\
		POMAP++&0:07:18&9& 0.0&0.0&0.0&0.0\\
		Holontology&0:05:18&9& 0.0&0.0&0.0&0.0\\
		DOME&3:49:07&9& 15688.7&0.61&0.61&0.61\\
		LogMap&3:54:43&7& 97081.4&0.08 (0.08)&0.14 (0.15)&0.81 (0.93)\\
		LogMapBio&0:39:00&9& 0.0&0.0&0.0&0.0\\ 
		LogMapLt&0:08:20&6& 82388.3&0.39 (0.39)&0.52 (0.56)&0.76 (0.96)\\
		Baseline&0:06:52&9& 17743.3&0.59&0.69 &0.82\\
		\hline
		\multicolumn{7}{|c|}{Overall performance}\\
		\hline
		AML&24:34:08&5&102471.1&0.19 (0.19)&0.23 (0.28)&0.31 (0.52)\\
		POMAP++&0:07:18&9& 16.9&0.79 &0.14 &0.08 \\
		Holontology&0:05:18&9& 18.8&0.80 &0.17 &0.10\\
		DOME&3:49:07&9& 15912.0&0.68 &0.68&0.67 \\
		LogMap&3:54:43&7& 97104.8&0.09 (0.09)&0.16 (0.16)&0.64 (0.74)\\
		LogMapBio&0:39:00&9& 24.1&0.68 &0.19 &0.11 \\
		LogMapLt&0:08:20&6& 88893.1&0.42 (0.42)&0.49 (0.54)&0.60 (0.77)\\
		Baseline&0:06:52&9& 17976.0&0.65 &0.73&0.82\\
		\hline
	\end{tabular}
	\label{tab:KGtrackResults2018}
\end{table}
\begin{table}[p!]
	\centering
	\caption{Knowledge graph track results for 2019, divided into class, property, instance, and overall correspondences. \cite{algergawy2019results}}
	\begin{tabular}{|l|c|c|c|c|c|c|}
		\hline
		System & Time & \# tasks & Size & Prec. & F-m. & Rec. \\
		\hline
		\multicolumn{7}{|c|}{Class performance}\\
        \hline
        AGM              & 10:47:38 & 5 & 14.6 & 0.23        & 0.09        & 0.06\\
		AML              &  0:45:46 & 4 & 27.5 & 0.78 (0.98) & 0.69 (0.86) & 0.61 (0.77)\\
		baselineAltLabel &  0:11:48 & 5 & 16.4 & 1.0         & 0.74        & 0.59\\
		baselineLabel    &  0:12:30 & 5 & 16.4 & 1.0         & 0.74        & 0.59\\
		DOME             &  1:05:26 & 4 & 22.5 & 0.74 (0.92) & 0.62 (0.77) & 0.53 (0.66)\\
		FCAMap-KG        &  1:14:49 & 5 & 18.6 & 1.0         & 0.82        & 0.70\\
		LogMap           &  0:15:43 & 5 & 26.0 & 0.95        & 0.84        & 0.76)\\
		LogMapBio        &  2:31:01	& 5 & 26.0 & 0.95        & 0.84        & 0.76)\\
		LogMapKG         &  2:26:14 & 5 & 26.0 & 0.95        & 0.84        & 0.76)\\
		LogMapLt         &  0:07:28 & 4 & 23.0 & 0.80 (1.0)  & 0.56 (0.70) & 0.43 (0.54)\\
		POMAP++          &  0:14:39 & 5 &  2.0 & 0.0         & 0.0         & 0.0  \\
		Wiktionary       &  0:20:14 & 5 & 21.4 & 1.0         & 0.8         & 0.67 \\
		\hline
		\multicolumn{7}{|c|}{Property performance}\\
		\hline
		AGM              & 10:47:38 & 5 & 49.4 & 0.66        & 0.32        & 0.21\\
		AML              &  0:45:46 & 4 & 58.2 & 0.72 (0.91) & 0.59 (0.73) & 0.49 (0.62)\\
		baselineAltLabel &  0:11:48 & 5 & 47.8 & 0.99        & 0.79        & 0.66\\
		baselineLabel    &  0:12:30 & 5 & 47.8 & 0.99        & 0.79        & 0.66\\
		DOME             &  1:05:26 & 4 & 75.5 & 0.79 (0.99) & 0.77 (0.96) & 0.75 (0.93)\\
		FCAMap-KG        &  1:14:49 & 5 & 69.0 & 1.0         & 0.98        & 0.96\\
		LogMap           &  0:15:43 & 5 & 0.0  & 0.0         & 0.0         & 0.0\\
		LogMapBio        &  2:31:01	& 5 & 0.0  & 0.0         & 0.0         & 0.0\\
		LogMapKG         &  2:26:14 & 5 & 0.0  & 0.0         & 0.0         & 0.0\\
		LogMapLt         &  0:07:28 & 4 & 0.0  & 0.0         & 0.0         & 0.0\\
		POMAP++          &  0:14:39 & 5 & 0.0  & 0.0         & 0.0         & 0.0\\
		Wiktionary       &  0:20:14 & 5 & 75.8 & 0.97        & 0.98        & 0.98 \\
		\hline
		\multicolumn{7}{|c|}{Instance performance}\\
		\hline
		AGM              & 10:47:38 & 5 & 5169.0 & 0.48        & 0.25        & 0.17\\
		AML              &  0:45:46 & 4 & 7529.8 & 0.72 (0.90) & 0.71 (0.88) & 0.69 (0.86)\\
		baselineAltLabel &  0:11:48 & 5 & 4674.2 & 0.89        & 0.84        & 0.80\\
		baselineLabel    &  0:12:30 & 5 & 3641.2 & 0.95        & 0.81        & 0.71\\
		DOME             &  1:05:26 & 4 & 4895.2 & 0.74 (0.92) & 0.70 (0.88) & 0.67 (0.84)\\
		FCAMap-KG        &  1:14:49 & 5 & 4530.6 & 0.90        & 0.84        & 0.79\\
		LogMap           &  0:15:43 & 5 & 0.0    & 0.0         & 0.0         & 0.0\\
		LogMapBio        &  2:31:01	& 5 & 0.0    & 0.0         & 0.0         & 0.0\\
		LogMapKG         &  2:26:14 & 5 & 29190.4& 0.40        & 0.54        & 0.86\\
		LogMapLt         &  0:07:28 & 4 & 6653.8 & 0.73 (0.91) & 0.67 (0.84) & 0.62 (0.78)\\
		POMAP++          &  0:14:39 & 5 & 0.0    & 0.0         & 0.0         & 0.0  \\
		Wiktionary       &  0:20:14 & 5 & 3483.6 & 0.91        & 0.79        & 0.70 \\
		\hline
		\multicolumn{7}{|c|}{Overall performance}\\
		\hline
		AGM              & 10:47:38 & 5 & 5233.2  & 0.48        & 0.25        & 0.17\\
		AML              &  0:45:46 & 4 & 7615.5  & 0.72 (0.90) & 0.70 (0.88) & 0.69 (0.86)\\
		baselineAltLabel &  0:11:48 & 5 & 4739.0  & 0.89        & 0.84        & 0.80\\
		baselineLabel    &  0:12:30 & 5 & 3706.0  & 0.95        & 0.81        & 0.71\\
		DOME             &  1:05:26 & 4 & 4994.8  & 0.74 (0.92) & 0.70 (0.88) & 0.67 (0.84)\\
		FCAMap-KG        &  1:14:49 & 5 & 4792.6  & 0.91        & 0.85        & 0.79\\
		LogMap           &  0:15:43 & 5 & 26.0    & 0.95        & 0.01        & 0.0\\
		LogMapBio        &  2:31:01	& 5 & 26.0    & 0.95        & 0.01        & 0.0\\
		LogMapKG         &  2:26:14 & 5 & 29216.4 & 0.40        & 0.54        & 0.84\\
		LogMapLt         &  0:07:28 & 4 & 6676.8  & 0.73 (0.91) & 0.66 (0.83) & 0.61 (0.76)\\
		POMAP++          &  0:14:39 & 5 &  19.4   & 0.0         & 0.0         & 0.0  \\
		Wiktionary       &  0:20:14 & 5 & 3581.8  & 0.91        & 0.8         & 0.71 \\
		\hline
	\end{tabular}
	\label{tab:KGtrackResults}
\end{table}
\begin{figure}[t]
	\centering
	\includegraphics[width=\textwidth]{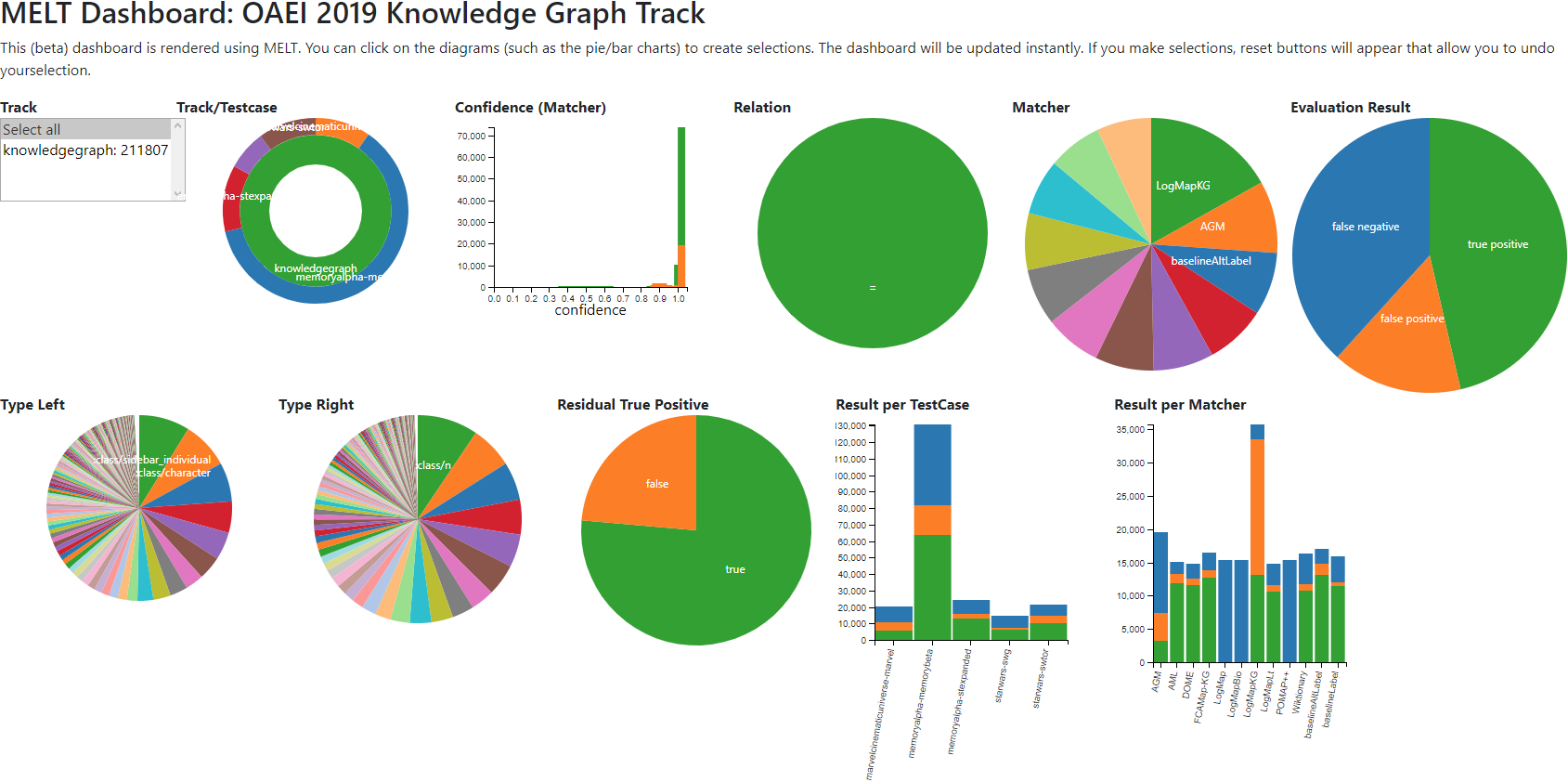}
	\caption{Dashboard for analyzing matcher results.}
	\label{fig:dashboard}
\end{figure}
Table \ref{tab:KGtrackResults} shows the aggregated results for all systems in 2019, including the number of tasks in which they were able to generate a non-empty alignment (\#tasks) and the average number of generated correspondences in those tasks (size). Like in the previous year, three systems (AML, DOME, and LogMapLt) were not able to solve all tasks. Again, the runtime differences are drastic, ranging from less than ten minutes to more than ten hours.

In addition to the global average precision, F-measure, and recall results, in which tasks where systems produced empty alignments were counted, we also computed F-measure and recall ignoring empty alignments which are shown in parentheses in the table, where applicable.

Nearly all systems were able to generate class correspondences. In terms of F-Measure, AML is the best one (when considering only completed test cases). Many matchers were also able to beat the baseline. The highest recall is about 0.77 which shows that some class correspondences are not easy to find.

In comparison to the 2018 edition, more matchers are able to produce property correspondences.
Only the systems of the LogMap family and POMAP++ do not return any alignments.
While Wiktionary and FCAMap-KG achieve an F-Measure of 0.98, other systems need more improvement here because they are not capable of beating the baseline (mostly due to low recall).

With respect to instance correspondences, AML and DOME are the best performing systems, but they outperform the baselines only by a small margin. On average, the systems returned between 3,000 and 8,000 instance alignments. Only LogMapKG returned nearly 30,000 mappings. The latter is interesting because LogMapKG is developed to prefer 1:1 alignments, but deviates here.
Thus, we conducted a deeper analysis of the alignment arity.
The results are shown in table \ref{tab:arity}. 
To account for matchers which return a mix of 1:n, n:1 or n:m mappings, not only the arity itself is reported, but also the count how often each appear in a mapping. For computing those numbers, the source of each correspondence is analyzed. If it links to only one concept, it counts as 1:1 if no other source is mapped to it, and otherwise as n:1. If the source links to multiple targets, it counts as 1:n and if one of those targets participate in multiple correspondences, the count for n:m is increased.

A strict 1:1 mapping is only returned by AGM, DOME and POMAP++, and the string matching baseline using only labels.
LogMap and LogMapBio return n:1 mappings in two test cases, whereas FCAMap-KG, Wiktionary, as well as the string matching baseline utilizing alternative labels, return a few n:m mappings in all test cases. AML and LogMapLt returned even more of those cases, and LogMapKG has the highest amount of n:m mappings. As discussed above, this is somewhat unexpected because the tool is tailored towards a track focusing only on 1:1 mappings.

\begin{table}[t!]
    \caption{Arity analysis of mappings produced in the knowledge graph track 2019.}
    \label{tab:arity}
    \centering
    \begin{tabularx}{\textwidth}{|X|r|r|r|r|r|r|}
         \hline
                 &       & mcu-               & memoryalpha- & memoryalpha- & starwars- & starwars- \\
         Matcher & arity & marvel                   & memorybeta   & stexpanded   & swg       & swtor \\
        \hline
        AMG              & 1:1 & 9,085 & 11,449 & 3,684 & 1,101 & 847 \\
        \hline
        \multirow{4}{*}{AML}
        & 1:1 & - & 14,935 & 3,568 & 3,323 & 3,755 \\
        & 1:n & - & 243    & 78    & 169   & 164 \\
        & n:1 & - & 3,424  & 281   & 74    & 103 \\
        & n:m & - & 240    & 69    & 12    & 24 \\
        \hline
        \multirow{4}{*}{baselineAltLabel}
        & 1:1 & 2,368 & 11,497 & 2,710 & 1,535 & 2,469  \\
        & 1:n & 54    & 855    & 277   & 114   & 131 \\
        & n:1 & 150   & 1,059  & 195   & 59    & 58 \\
        & n:m & 2     & 103    & 48    & 4     & 7 \\
        \hline
        baselineLabel & 1:1 & 1,879 & 10,552 & 2,582 & 1,245 & 2,272 \\
        \hline
        DOME & 1:1 & - & 12,475 & 2,727 & 2,024 & 2,753 \\
        \hline
        \multirow{4}{*}{FCAMap-KG}
        & 1:1 & 2,510 & 12,423 & 2,985 & 1,828 & 2,620  \\
        & 1:n & 28    & 288    & 94    & 240   & 125 \\
        & n:1 & 138   & 382    & 76    & 47    & 37 \\
        & n:m & 6     & 78     & 19    & 25    & 14 \\
        \hline
        \multirow{2}{*}{LogMap}
        & 1:1 & 12 & 32 & 33 & 14 & 29  \\
        & n:1 & 0   & 8 & 0  & 0  & 2 \\
        \hline
        \multirow{2}{*}{LogMapBio}
        & 1:1 & 12 & 32 & 33 & 14 & 29  \\
        & n:1 & 0   & 8 & 0  & 0  & 2 \\
        \hline
        \multirow{4}{*}{LogMapKG}
        & 1:1 & 2,919  & 10,453 & 2,600 & 1,663 & 2,122  \\
        & 1:n & 1,363  & 4,741  & 2,857 & 6,596 & 7,797 \\
        & n:1 & 3,207  & 2,963  & 1,016 & 410   & 218 \\
        & n:m & 33,593 & 36,382 & 9,089 & 6,668 & 9,425 \\
        \hline
        \multirow{4}{*}{LogMapLt}
        & 1:1 & - & 12,935 & 3,349 & 2,500 & 3,217  \\
        & 1:n & - & 270    & 119   & 205   & 293 \\
        & n:1 & - & 2,881  & 36    & 50    & 95 \\
        & n:m & - & 602    & 73    & 52    & 30 \\
        \hline
        POMAP++ & 1:1 & 9 & 20 & 25 & 14 & 29 \\
        \hline
        \multirow{4}{*}{Wiktionary}
        & 1:1 & 1,757 & 9,274 & 1,975 & 1,494 & 2,321  \\
        & 1:n & 26    & 246   & 110   & 72    & 104 \\
        & n:1 & 74    & 162   & 58    & 18    & 14 \\
        & n:m & 8     & 156   & 24    & 8     & 8 \\
        \hline
    \end{tabularx}
\end{table}

For a further detailed analysis of the track results, an online dashboard\footnote{\url{http://oaei.ontologymatching.org/2019/results/knowledgegraph/knowledge_graph_dashboard.html}} is implemented. The user interface is shown in figure \ref{fig:dashboard}. It is mainly intended for matcher developers to analyze their results and improve their systems. The basis is a table of all correspondences together with the evaluation result. The charts at the top allow a filtering of these correspondences by different criteria which can also be combined. The code for generating the dashboard is included in the MELT framework~\cite{melt} to enable system developers to generate their own analyses.

Some of the key observations of the 2019 edition of the knowledge graph track include:
\begin{enumerate}
    \item There is no one-size-fits-all solution. Instead, we can observe that different matchers produce the best results for classes, properties, and instances.
    \item Scalability is an issue, since not all matchers are capable of solving all tracks, and the runtime varies drastically between the different systems.
\end{enumerate}

\subsection{Hidden Task in OAEI 2019 and the Golden Hammer Bias}
So far, we have only analyzed settings in which matchers were provided with two knowledge graphs from the same domain. In other words: it is already known that some correspondences are to be found. This is the usual setup in OAEI tracks, where correspondences between the input ontologies are always expected.

In many real world scenarios, we cannot make that assumption. We refer to those scenarios as \emph{open} scenarios, in contrast to \emph{closed domain} scenarios, where the input ontologies share a domain. All OAEI tracks evaluate the latter kind, i.e., using ontologies from the conference or medical domain etc. In contrast, the matching in DBkWik, where thousands of knowledge graphs from different domains are to be integrated, is an open scenario. In such a scenario, where thousands of knowledge graphs co-exist, a random pair of two knowledge graphs may or may not have a certain share of entities in common.

In order to find out whether tools are over-tuned towards closed-domain scenarios, we introduced a \emph{hidden track} to the 2019 edition, i.e., an evaluation which we did not inform the participants about. For this track, we used the single graph within the DBkWik set with the largest number of instances -- i.e., the one extracted from \emph{LyricWiki}\footnote{\url{https://lyrics.fandom.com/}}, which has more than 1.7M instances (we took a sample of about one million instances to reduce the runtime of the matchers). Since the main classes are songs, albums, and music artists, we expect a rather low overlap with the other graphs in the KG track, which come from different domains. At the same time, we expect a high overlap of trivial string matches, since there are songs, albums, or artists called \emph{Star Trek}, \emph{Star Wars}, etc., contained in the KG.

\begin{figure}[t]
	\centering
	\includegraphics[width=\textwidth]{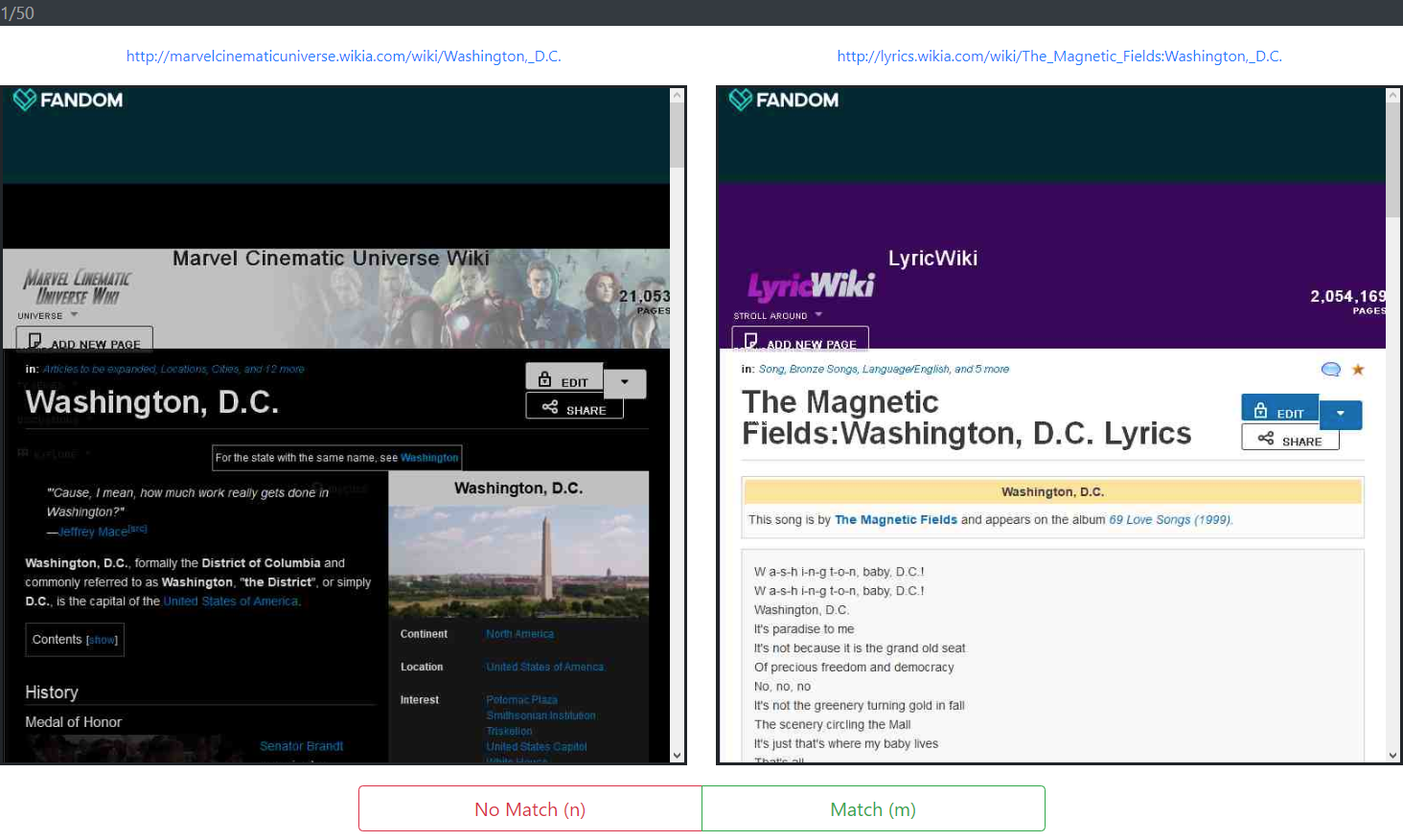}
	\caption{User interface for judging if two Wiki pages correspond to the same concept. }
	\label{fig:crowdSourcingInterface2}
\end{figure}

\begin{table}[t]
    \caption{Test cases with lyric wiki as target. For each matcher and test case 50 correspondences were analyzed.}
    \label{tab:lyricscomparison}
    \centering
    \begin{tabularx}{\textwidth}{|X|r|r|r|r|r|r|}
         \hline
         & \multicolumn{2}{c|}{mcu} & \multicolumn{2}{c|}{memoryalpha} & \multicolumn{2}{c|}{starwars} \\
         & \multicolumn{2}{c|}{lyrics} & \multicolumn{2}{c|}{lyrics} & \multicolumn{2}{c|}{lyrics} \\
         \hline
         Matcher & matches & precision & matches & precision & matches & precision \\
        \hline
        AML              & 2,642   & 0.12 & 7,691 & 0.00 & 3,417 & 0.00 \\
        baselineAltLabel & 588     & 0.44 & 1,332 & 0.02 & 1,582 & 0.04 \\
        baselineLabel    & 513     & 0.54 & 1,006 & 0.06 & 1,141 & 0.06 \\
        FCAMap-KG        & 755     & 0.40 & 2,039 & 0.14 & 2,520 & 0.02 \\
        LogMapKG         & 29,238  & 0.02 & -     & -    & -     & - \\
        LogMapLt         & 2,407   & 0.08 & 7,199 & 0.00 & 2,728 & 0.04 \\
        Wiktionary       & 971     & 0.12 & 3,457 & 0.02 & 4,026 & 0.00 \\
        \hline
    \end{tabularx}
\end{table}
All matchers which participated in the knowledge graph track in 2019 were executed on three test cases. Those test cases always have the lyrics Wiki has the target and the following three Wikis as a source: Marvel Cinematic Universe (mcu), Memory Alpha and Star Wars. Since we cannot rule out true positives completely, we evaluated the precision manually by sampling 50 correspondences from the result sets for each matcher and testcase, which totals more than 1k samples (7 matchers x 3 test cases x 50 samples). With this sample size the maximum error is 15\% at a 0.95 confidence level. A web front end depicted in figure~\ref{fig:crowdSourcingInterface2} is developed to help the annotators judging if two concepts are the same. It shows two images of the corresponding Wiki page (which are created with phantomjs\footnote{\url{https://phantomjs.org/}}) to provide a constant, browser-independent visualization, and to prevent clicking on links. 

Not all matchers are able to execute this hidden task. LogMap, LogMapBio and POMAP++ find only schema mappings and DOME needs more than the provided memory (even when provided with 100 GB of RAM). AGM throws a tool exception, and LogMapKG is only able to finish one test case (in all other cases it runs into a 24 hour timeout). These findings, involving a larger knowledge graph (despite smaller than, e.g., DBpedia or Wikidata) illustrate that scalability is still an issue.

The results are shown in table~\ref{tab:lyricscomparison}. 
We can see that all matchers find a considerable amount of instance matches, on average more than one thousand per pair of knowledge graphs. At the same time, the precision is really low. If we contrast those precision figures with the ones in table~\ref{tab:KGtrackResults}, we can observe some crucial differences. For all matchers, including the baselines, the precision figures in the latter are significantly higher than those in the hidden track.

This illustrates that all tools make an implicit assumption that some overlap between the ontologies provided exists, and create a certain amount of nonsensical results if that assumption is not met. We can observe this very drastically in the case of matching memoryalpha to lyrics, where the tools match between 2\% and 17\% of the instances in the smaller Wiki, with most of those matchings being false positives.

We refer to this observation as the \emph{golden hammer bias}: in evaluation setups such as the OAEI (as well as most other evaluations of matching tools), the performance of the matching tools is systematically over-estimated. In contrast, when applying a tool in an open scenario where a positive outcome is not guaranteed a priori, the approaches at hand assume such a positive outcome nevertheless, and, hence, cannot solve the task properly. In particular, this is the case for LogMapKG, which creates a very large number of mappings at a very low precision, at least for the task it is able to solve.

The case of mcu is particularly interesting, since a significant portion of true matches can actually be found here (e.g., songs used in movies). Nevertheless, the precision of all tools is much lower than the precision on tasks in pure closed domain scenarios.

As a conclusion, we can see that existing matching tools cannot be used in open domain scenarios out of the box. Further filtering or a priori class-wise or even knowledge-graph wise blocking would be necessary, although, in the latter case, a significant amount of true positives, like in the case of mcu, would be missed.

\section{Conclusion and Outlook}
\label{sec:conclusion}
In this paper, we have described the design of knowledge graph track at the OAEI which focuses on the simultaneous matching of instances and schemas. 
We have introduced the datasets based on the DBkWik knowledge graph extracted from thousands of Wikis, and we have discussed two different strategies for creating the gold standard for instance links -- i.e., by crowdsourcing and by utilizing explicit interlinks between Wikis. 
Moreover, we have introduced a hidden track to inspect the effect of tools expecting a positive outcome of a task, which we named the \emph{golden hammer bias}.

From the results in 2018 and 2019, we can make different observations. 
First, the task is inherently hard, with most of the tools outperforming a simple string matching baseline only by a small margin. 
Second, there are strategies for individual subtasks -- i.e., matching classes, properties, and instances -- which clearly outperform others, including the baseline, but no tool implements a strategy that consistently outperforms the others.
Third, many tools have difficulties handling larger-scale input data, i.e., knowledge graphs with millions of instances.

An additional evaluation using a hidden track revealed yet another issue with current knowledge graph matching tools. 
All of them make the tacit assumptions that the knowledge graphs to be matched have something in common. 
When confronted with two unrelated knowledge graphs, we have shown that they produce thousands of mostly false matches. 
We call this effect the \emph{golden hammer bias} -- the tools are applied without considering whether they are actually applicable.

Since we observe a growing number of tools that use supervised methods for matching knowledge graphs, we plan to create a sub-track which supports this setting, e.g., by providing separate training, testing, and validation sets.

In sum, those findings show that the task of knowledge graph matching is far from being solved.
With our ongoing evaluation efforts -- the KG track will be part of OAEI 2020 again -- we provide a testbed for new and improved solutions in that field. 
Moreover, a few approaches for the task of knowledge graph matching have been published in the recent past (e.g., \cite{li2019semi,sun2018bootstrapping,trisedya2019entity}), which have been evaluated on different datasets (and in closed domain settings only), hence, their results are not directly comparable.
By providing a generic benchmark including both open and closed domain settings, we enable a more systematic comparison for such works in the future.

\bibliographystyle{splncs04}
\bibliography{bibliography}

\end{document}